\documentstyle[11pt,newpasp,twoside,epsf]{article}
\begin{document}
\title{Helical structures in Seyfert galaxies}
\author{ A.V. Moiseev, V.L. Afanasiev, S.N. Dodonov}
 \affil{ Special Astrophysical Observatory, Nizhnij Arkhyz,
 Karachaevo-Cherkesia, 369167, Russia }
\author{V.V. Mustsevoi  and S.S. Khrapov}
 \affil{Volgograd State University, Volgograd, 400062, Russia}

\begin{abstract}
The Seyfert galaxies with Z-shaped   emission
filaments    in the  Narrow Line  Region (NLR)   are considered. We  assume 
that observable Z-shaped structures and velocity pattern of NLR may
be explained as  tridimensional helical waves in the ionization
cone. 
\end{abstract}

\section{Introduction}
The numerous emission line images  of  Seyfert galaxies 
show a existence of a cone-like  NLR with  a broad opening angles 
and spatial sizes from $10$~pc  to $18$ kpc (Wilson \& Tsvetanov 1994). Also 
these galaxies has high-collimated elongated radio structures
(radio-jets)  coinciding with the cones axis (Wilson \&
Tsvetanov 1994; Nagar et al. 1999). A ordered Z-shaped emission pattern  is a 
frequently  features in the NLRs.  We found more then 20 such
objects  on  published emission-line images of the nearby Seyferts.
There are no   common point of view on an origin  of the 
such regular structures  in  NLR. Different models were proposed for
individual objects:  a  bent bipolar outflow (Mulchaey et al.
1992), strong  collimated precessing twin-jet (Veilleux, Tully, \& 
Bland-Hawtorn 1993), a system of inclined gaseous disks (Morse et al. 1998) 
and etc.

\section{Kinematical features of the Z-shaped NLR}

A sample of  galaxies  with Z-shaped NLRs were  observed at the 6m
telescope. The scanning Interferometer Fabry-Perot and integral field 
spectrograph MPFS were used for study of the 2D kinematics of stars and 
ionized gas. Some systems of the gas clouds with   velocity 
difference more than $100\div 200\,\mbox{km}\,\mbox{s}^{-1}$ are present on 
the light of sight in the central region of cones,  but the outer emission 
filaments has only one component of the emission lines. The  gas velocity 
fields  are strongly non-circular in comparing with the stellar one. 
A  large gradient of the line of sight gas velocities
presents in  Z-shaped spiral. We note that similar features also observed on 
the velocity  fields in the other Sy galaxies with (see references in Moiseev 
et al. 2000)  and these could be   common for Z-shaped pattern.

\section{Non-linear simulations}

We suggest that Z-shaped spiral  filaments in NLRs have a common wave  origin 
and generated by  the hydrodynamic instability  due to the velocity
break between a galactic instellar medium and a outflowing gas from the central
AGN. A collimated radio jet  corresponds to the direction
of outflow and matches with the cone's axis.
A linear  analysis shows that jet are 
unstable relatively to waveguide-resonance development 
 of pinch and helical internal gravity waves. A developments of the
instabilities  leads to a set of shock waves creation in an
ambient medium. Our 2D and 3D non-linear hydrodynamical simulations   
show that  the shock waves penetrate outward from the jet boundaries to the 
ambient medium. A resulting shock-wave  structure  covers  broad cone 
with open angle  $30^\circ  \div 80^\circ$ and   
appears in a sky-plane
as a NLR with bright emission pattern (Fig.~1).
A pinch (axisymmetrical) and helical modes of the shock waves
are develop in the NLR. A helical wave modes provides the
Z-shaped  emission  structures  which  observed  in  the ionized
cones. 

\acknowledgements{
This work was supported by federal program "Astronomy" (Project 1.2.3.1)}

\begin{figure}
\plotone{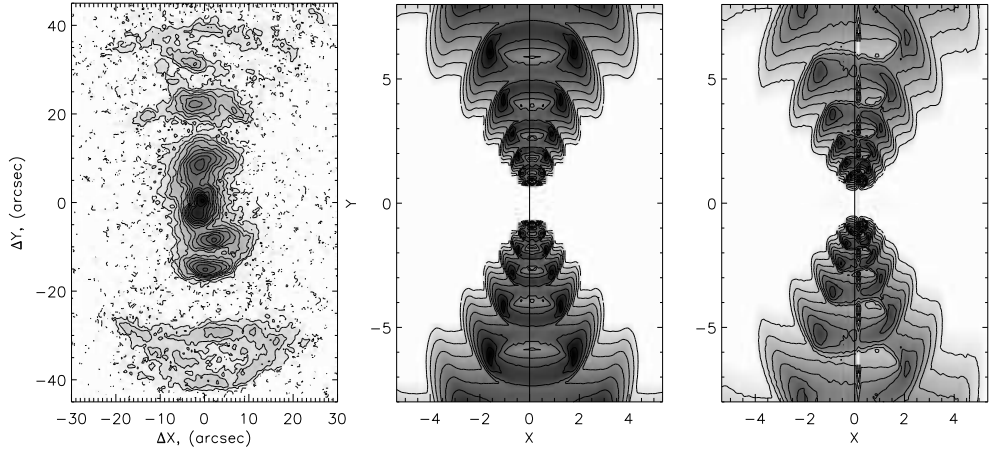}
\caption{ The  [OIII] image of NGC 5252 obtained at the 6m telescope ({\it  
left}) and contours  of  the model luminosity for pinch  ({\it  middle}) 
and for helical ({\it right}) shock wave modes.
}
\end{figure}

\end{document}